\documentclass[referee]{aa}
\usepackage{graphicx,amssymb,epsf}
\newcommand{\AJ}{AJ}
\newcommand{\ARAA}{ARA\&A}

\newcommand{\AaA}{A\&A}

\newcommand{\ApJ}{ApJ}
\newcommand{\MNRAS}{MNRAS}
\newcommand{\Natur}{Nature}

\begin{document}
\title{How accurately can the SZ effect measure peculiar cluster 
velocities and bulk flows?}
\author{N. Aghanim\inst{1} \and K.M. G\'orski\inst{2}\inst{,3} \and J.-L. Puget\inst{1}}
\institute{IAS-CNRS, B\^atiment 121, Universit\'e Paris Sud, F-91405 Orsay 
Cedex, France 
\and E.S.O., Karl-Schwarzschild-Str. 2, D-85748 Garching bei M\"{u}nchen,
Germany \and Warsaw University Observatory, Poland}
\date{Received date / accepted date}
\abstract{
The Sunyaev-Zel'dovich effect is a powerful tool for cosmology that
can be used to measure the radial peculiar velocities of galaxy
clusters, and thus to test, and constrain, theories of structure 
formation and evolution.
This requires, in principle, an accurate measurement of
the effect, a good separation between the Sunyaev-Zel'dovich
components, and a good understanding of the sources contributing to the 
signal and
their effect on the measured velocity. In this study, we evaluate the
error in the individual radial peculiar velocities determined with 
Sunyaev-Zel'dovich measurements. We estimate, for 
three cosmological models, the errors induced by the
major contributing signals (primary Cosmic Microwave Background anisotropies, 
Sunyaev-Zel'dovich effect due to the background cluster population,
residuals from component separation and instrumental noise).  We generalise 
our results to estimate the error in the bulk velocity on large scales. In
this context, we
investigate the limitation due to the Sunyaev-Zel'dovich 
source (or spatial) confusion in a Planck-like instrumental configuration. 
Finally, we propose a strategy based on the future all-sky 
Sunyaev-Zel'dovich survey, that will be provided by the Planck mission, to
measure accurately the bulk velocities on large scales up to redshift 1, or 
more.
}
\maketitle
\markboth{How accurately can the SZ effect measure peculiar
and bulk velocities?}{}
\section{Introduction}
One of the major fields of research in cosmology is the study of the
large scale matter distribution in the universe. On large
scales, the evolution of the matter density fluctuations is adequately
described
through linear physics. The matter distribution thus represents the
imprint of the initial density perturbations. Combined with other
results, such as those derived from Cosmic Microwave Background (CMB)
observations, this kind of information can probe the structure
formation models and the cosmological parameters. \par 
The luminous matter distribution can be probed through the
direct observation of galaxy, or galaxy cluster, distributions, but this
gives a biased view of the total matter distribution. The latter
can be inferred from the velocity fields. In fact, the
inhomogeneities in the mass distribution produce deviations from the
Hubble flow referred to as the peculiar velocities. A few methods have been
suggested to measure the transverse velocity components, e.g.,
\cite{birkinshaw83a} and \cite{birkinshaw83}. However, in practice, 
the radial component of the peculiar velocities can only be measured using
redshift surveys, and by some established relationships which give independent 
distances to the objects, such as the Faber-Jackson \cite[]{faber76} or the
Tully-Fisher \cite[]{tully77} relations. Nevertheless, these methods lead 
to uncertainties in the expansion velocity which are proportional to the 
distance and
thus induce even larger uncertainties when measuring peculiar velocities 
on very large scales. The velocity fields can be 
studied using galaxies 
or galaxy clusters. However, there are more advantages in studying the
deviations from the Hubble flow as traced by galaxy clusters.  One of
these advantages comes from the fact that, on scales probed by galaxy 
clusters,
the underlying density fluctuations are largely in the linear regime
and therefore very close to the initial conditions from which large
scale structures developed. Finally, given observed radial peculiar 
velocities and the potential flow assumption, the full
three-dimensional velocity field can be derived using 
reconstruction methods (\cite[]{bertschinger89,zaroubi99} and references 
therein). The
inferred density field is, in principle, representative of both the
dynamical evolution of the structure and the underlying total
matter distribution. Therefore, comparisons between the reconstructed
total matter distribution from velocity fields and CMB fluctuations 
and that traced by galaxies can test and constrain
theories of structure formation and evolution, and cosmological
models.  \par\medskip
Moreover, given cluster radial peculiar velocities at known positions, cluster 
motions can be used as cosmological probes using various statistical  
quantities among which are the velocity frequency distribution, the
power spectrum, the velocity dispersions, the pairwise velocities, the
velocity correlation function, and the bulk flows. As already mentioned, 
relative errors in the 
distance determination lead to the peculiar velocity uncertainty increasing
linearly with distance. As a consequence, all
the statistical quantities derived from observed velocities suffer
from large errors especially at high redshifts. For example, in the case of 
the bulk
velocity, which is the average of the local velocity field smoothed
over a window function of scale $Rh^{-1}$ Mpc ($h=H_0/(100\, 
\mbox{km/s/Mpc})$ is the normalised Hubble constant), several measurements
exist in literature
\cite[]{dressler87,courteau93,willick96,giovanelli98,hudson99,willick99}.
Whereas most of these studies agree on the reality of a significant
bulk flow within $50\,h^{-1}$ Mpc; the situation is more controversial
at larger scales where the accuracy of the determination of the bulk 
velocities decreases.
\par\bigskip 
In order to overcome the problem of large uncertainties
in the peculiar velocities, we must find ways of measuring them with a
redshift independent accuracy. In this context, the
Sunyaev-Zel'dovich (SZ) effect is a very promising, and potentially
sensitive, tool. As proposed
initially by \cite{suniaev80}, the radial peculiar velocity of galaxy
clusters can be determined using SZ measurements, which are in addition
distance independent. In the forthcoming years several experiments (ground 
based, balloon borne or space) will measure the SZ effect and
perform SZ surveys (AMIBA, ARCHEOPS, BOLOCAM, MAP, ...). These 
experiments, of which the
Planck satellite\footnote{http://astro.estec.esa.nl/Planck}
is the best example, are designed so that the sensitivity, the
frequency coverage and the angular resolution allow a very good
separation of the two SZ effect components (thermal and kinetic, see
Sect. \ref{sec:sz}), and a best possible evaluation of the foreground 
(galactic
and extra-galactic) emissions that contribute to the measured signal. In
this context, we expect that the number of observed SZ clusters will
rapidly increase thus allowing the radial peculiar velocity of clusters to be 
measured and providing a useful cosmological tool. This kind of project
has already been undertaken on a sample of 40 known galaxy
clusters with redshifts ranging between 0.1 and 0.3 by the SUZIE team
\cite[]{holzapfel97}. 
Nevertheless, the accuracy in individual clusters is limited by the 
contamination from other components of the microwave sky. It has been suggested
by \cite{aghanim97} and \cite{kashlinsky2000} that averaging over many
clusters in a large volume is probably the best method to measure the 
very large scale velocity field up to $z=1$. This requires a sensitive all-sky
SZ survey which will be provided only by the Planck 
mission. In this paper, we investigate quantitatively the accuracy of such
a measurement, within the context of Planck.
\par\bigskip 
As described in sections 2 and 3, the combination of the
so-called kinetic and thermal SZ effects allows us to measure the radial
peculiar velocity. However, this measurement is affected by errors
due to astrophysical contributions (CMB, background
cluster population), to residuals from the component separation and to
instrumental noise. In section 5, we present the method used to
evaluate the {\it rms} error which affects the peculiar velocity
measurement.  We give, for three cosmological models, the results obtained
for the major sources of error in terms of the {\it rms} error as a
function of the cluster size for a Planck measurement. These errors are 
evaluated using 
simulated maps. The cluster model and the map simulation are presented
in section 4.  In section 6, we generalise the computation of the {\it rms} 
error to the bulk velocities at an illustrative scale of 100$h^{-1}$ Mpc. We
discuss our results and conclude in section 7. \\ 
In the following, we use the baryon density
$\Omega_b=0.06$ \cite[]{walker91}. We discuss three cosmological
models: an open model with a density parameter $\Omega_m=0.3$, and 
two flat models: one with a non-zero cosmological constant
($\Omega_m=0.3$ and $\Omega_\Lambda=0.7$), and the other with no
cosmological constant ($\Omega_m=1$).
\section{The SZ (thermal and kinetic) effect} \label{sec:sz}
The SZ effect comprises the so-called thermal and
kinetic effects. The thermal SZ effect is the inverse Compton
interaction between CMB photons and the
free electrons of the hot intra-cluster medium. Its amplitude is
characterised by the Compton parameter $y$ -- the integral of
the pressure along the line of sight -- which depends only on the
cluster electron temperature and density ($T_e$, $n_e$):
\begin{equation} 
y=\frac{k\sigma_T}{m_ec^2}\int T_e(l)n_e(l)\,dl\, ,
\label{eq:yparam}
\end{equation} 
\noindent
where $k$ is the Boltzmann constant, $\sigma_T$ the Thomson cross section,
$m_e$ the electron mass, $c$ the speed of light and $l$ is the distance 
along the line of sight. If the intra-cluster gas is isothermal 
($T_e(l)=T_e=\mbox{const}$), $y$ is
expressed as a function of the optical depth $\tau$ ($\tau=\sigma_T\int n_e(l)
\,dl$): 
\begin{equation}
y=\tau~\frac{kT_e}{m_ec^2}.
\label{eq:yt}
\end{equation}
The inverse Compton interaction conserves the number of photons and shifts 
their spectrum, on average, to higher frequencies. This can be observed 
as the induced relative monochromatic intensity difference between Compton
distorted and undistorted CMB:
\[
\frac{\Delta I_{\nu}}{I_{\nu}}=y\cdot f(x),
\]
\noindent
where $x$ is the dimensionless frequency $x=h_{pl}\nu/kT_{CMB}$ 
($h_{pl}$ denotes the
Planck constant, $T_{CMB}$ the CMB temperature, and $\nu$ the frequency),
$I_{\nu}$ is the intensity of the CMB (black body emission) and $f(x)$ is the
spectral form factor given by:
\[
f(x)=\frac{xe^x}{(e^x-1)}\left[x\left(\frac{e^x+1}{e^x-1}\right)-4\right].
\]
In the non-relativistic approximation, this spectral signature is universal,
whereas it varies with the temperature of the intra-cluster gas in the exact
computations including the relativistic corrections
\cite[]{wright79,rephaeli95,challinor98,itoh98,pointecouteau98,sazonov98,hansen99,molnar99,nozawa2000}. 

When the cluster moves with a radial peculiar velocity $v_r$, an additional 
relative intensity variation of the CMB due to the first-order Doppler effect
is generated. It is: 
\[
\frac{\Delta I_{\nu}}{I_{\nu}}=- \frac{v_r}{c}\tau\cdot a(x)=
\left(\frac{\Delta T}{T}\right)_{SZ}\cdot a(x), 
\]
where $a(x)$ is a spectral form factor, given by:
\[
a(x)=x\frac{e^x}{e^x-1}.
\]
This is what is commonly referred to as the kinetic SZ effect. The
intensity fluctuation induced by the kinetic SZ effect has the same
spectral shape as the primordial anisotropies (equivalent to a temperature
fluctuation). The amplitude of the temperature anisotropy induced by
the kinetic SZ effect is thus:
\begin{equation} 
\left(\frac{\Delta T}{T}\right)_{SZ}=-\frac{v_r}{c}\,\tau.
\label{eq:dt}
\end{equation} 
The effect is positive for clusters moving
towards the observer (i.e. with negative velocities).  \par
\section{Measuring the peculiar velocity with SZ}
The characteristic spectral signature of the thermal SZ effect
makes it a powerful tool for detecting galaxy clusters through
millimetre and submillimetre observations. The kinetic SZ effect 
has a very different spectral signature from the thermal effect. In
particular, it peaks at about 1.4 mm where the thermal effect is null.
A multifrequency observation of a galaxy cluster should therefore 
allow a separation of the two effects. Consequently, the SZ effect can in
principle be used as a tool for measuring the radial component of the
galaxy cluster peculiar velocity $v_r$. In fact, when we combine the
thermal and kinetic SZ (Eqs. \ref{eq:yt} and \ref{eq:dt}), we obtain:
\begin{equation}
v_r=-c\,\frac{kT_e}{m_ec^2}\frac{(\delta T/T)_{SZ}}{y}.
\label{eq:vr}
\end{equation}
This method was first suggested by \cite{suniaev80}, who also proposed
to use it for bulk motion measurements at large scales. \cite{rephaeli91} 
made one of the first estimates of the possibility of measuring the peculiar
velocities using a selected sample of galaxy clusters. However, the most 
convincing measurements on individual clusters have only been done 
recently with the new generation of very sensitive bolometers
\cite[]{holzapfel97,lamarre98}. 
\par\bigskip 
As mentioned in the previous section, the primary CMB
anisotropies have the same spectral signature as the kinetic SZ effect
of galaxy clusters. Therefore, the measured $(\delta T/T)$ towards a 
targeted cluster is 
contaminated by the primordial temperature fluctuations of the
CMB. It is also contaminated by the background fluctuations induced by the 
clusters population, by any non-removed or residual astrophysical foreground
contribution, and finally by the instrumental noise. These sources of
contamination are responsible for an 
error $\delta v_r$ in the estimated cluster peculiar velocity. When 
computed using Eq. (\ref{eq:vr}), the relative error in the velocity 
can be expressed as follows:
\begin{equation}
\frac{\delta v_r}{v_r}=\frac{\delta A}{A} + \frac{\delta T_e}{T_e}, 
\label{eq:dvr}
\end{equation} 
where $A=\frac{(\delta T/T)}{y}$. The $\delta T_e/T_e$ term is the relative
error due to the uncertainty in the intra-cluster gas temperature
which should be derived from X-ray data or from the SZ
effect itself, as proposed by \cite{pointecouteau98}. For the new generation
of X-ray satellites (Chandra, XMM-Newton), this error is expected 
to be of the order of 5 to 10\%. Hereafter, we will neglect
this source of error in the evaluation of 
$\delta v_r$. The error due to the CMB primary fluctuations enters into 
the term $A$ of Eq. (\ref{eq:dvr}) through the measurement of $\delta
T/T$. In the same manner, the fluctuations generated by the background 
population of galaxy clusters through their kinetic SZ effect and all
spurious emissions (astrophysical residuals, noise, ...) will also contribute
to the error in $\delta v_r$ through the $\delta T/T$ term. The residuals 
of the component
separation between thermal and kinetic SZ effects should contribute either
to the $\delta T/T$ term, or to the measurement of the cluster Compton 
parameter $y$. In our study, we take them into account as an additional 
$\delta T/T$ component. \par
Based on a map analysis, we evaluate the error in the peculiar velocity of 
individual 
galaxy clusters due to all contributing sources (CMB, clusters, ...)  
and we express it in terms of an induced {\it rms} error in the velocity. 
In this context,
each source of error, $i$, will contribute in a quadratic form to the 
overall {\it rms} error, that is 
$(\delta v_{rms}^{tot})^2=\sum_i (\delta v_{rms}^i)^2$. %
\section{Simulating the SZ effect of galaxy clusters} \label{sec:simu}
We use a set of simulated maps for both the primary CMB fluctuations and the
SZ effect contribution due to the thermal and the kinetic effects. 
All maps have $512\times 512$ 1.5 arcmin square pixels. 
We simulate the SZ effect using an empirical approach which consists of
predicting the number of galaxy clusters that were formed between today and
a redshift $z=10$. To do so, we use the Press-Schechter (PS) mass function 
\cite[]{press74}. The individual galaxy clusters are modelled following a
$\beta$-profile (see Sect. \ref{sec:mod}) and their positions on the
simulated maps are drawn at random. 
\subsection{Cluster counts}
The general analytic expression for the PS counts gives the comoving 
number density of 
spherical collapsed halos in the mass range $[M,M+dM]$ formed at a 
redshift $z$:
\begin{equation}
\frac{dn(M,z)}{dM}=-\,\sqrt{\frac{2}{\pi}}\,\frac{\overline{\rho}}{M^2}
\,\frac{d\ln\,\sigma(M,z)}{d\ln\,M}\,\frac{\delta_{c0}(z)}{\sigma(M,z)}\,
\exp\left[-
\frac{\delta_{c0}^2(z)}{2\sigma^2(M,z)}\right],
\label{eq:pscount}
\end{equation} 
where $\overline{\rho}$ is the mean comoving background density 
and $\delta_{c0}(z)$ is the overdensity of a linearly evolving
structure.  The mass variance $\sigma^2(M,z)$ of the fluctuation
spectrum, filtered on mass scale $M$, is related to the power
spectrum of the initial density fluctuations $P(k)$ \cite[]{peebles80}. 
Following 
\cite{viana96} and \cite{viana99}, we use an approximation of the variance
in spheres of radius $R$ ($R=(3M/4\pi\overline{\rho})^{1/3}$) in
the vicinity of 8$h^{-1}$ Mpc:
\[
\sigma(R,z)=\sigma_8(z)\left(\frac{R}{8h^{-1}\,\mbox{Mpc}}\right)^{-\gamma(R)},
\]
with
\[
\gamma(R)=(0.3\Gamma+0.2)\left[2.92+log\left(\frac{R}{8h^{-1}\,\mbox{Mpc}}\right)\right].
\]
$\Gamma$ is the so-called shape parameter of the cold dark matter transfer
function, taken to be 0.23 (see \cite{viana99} for a discussion).
The redshift evolution $\sigma_8(z)$ is given by the perturbation growth law
\cite[]{carroll92}.
\subsection{Modeling individual clusters} \label{sec:mod}
The spatial distribution
of the SZ (thermal and kinetic) effect is ruled by the intra-cluster
gas profile. The latter is generally well-described by the so-called 
$\beta$-profile \cite[]{king66}. We will therefore use for simplicity, as in 
\cite{cavaliere78}, the hydrostatic isothermal model with a spherical geometry.
In this model the electron density distribution is given by:
\begin{equation}
n_e(R)=n_{e0}\left[1+\left(\frac{R}{R_c}\right)^2\right]^{-\frac {3\beta
}{2}},
\end{equation}
where $n_{e0}$ is the central electron density, $R_c$ is the cluster core 
radius cluster and $\beta$ is a parameter whose value is about 2/3 as 
indicated 
by both numerical simulations \cite[]{evrard90} and X-ray surface brightness 
profiles \cite[]{jones84,edge91}. The physical parameters of a galaxy cluster
(temperature, virial radius and central electron density) can be computed once
its formation redshift and its mass are known. 
The cluster temperature $T_e$, in keV, is given by \cite{bryan98}:
\begin{equation}
T_e=1.39\frac{f_T}{b}\,M_{15}^{2/3}\,[h^2\Delta_c(z)E(z)^2]^{1/3},
\label{eq:te}
\end{equation}
where $f_T$ and $b$ are numerical factors set respectively to 0.79 and 1,
$M_{15}$ is the cluster mass in $10^{15}\,M_\odot$ units, $\Delta_c(z)$
is the critical density (expressions can be found in \cite{bryan98}) and 
$E(z)$ is related to the time by
$t(z)=H_0^{-1}\int_z^\infty\,(1+z)^{-1}E(z)^{-1}$ (see \cite{peebles93} for 
example). \par
The central density $n_{e0}$ can be derived from the cluster gas mass using 
the following equation:
\begin{equation}
M_{gas}\left(\frac{\Omega_b}{\Omega_m}\right)=m_p\mu\int_0^{R_{vir}}n_e(R)
\,4\pi R^2\,dR, 
\end{equation}
where the virial radius of the structure, for a critical universe 
($\Omega_m=1$), is given by: 
\begin{equation}
R_{vir}=\frac{(G\,M)^{1/3}}{(3\pi\,H_0)^{2/3}}\,\frac{1}{1+z},
\end{equation}
In this equation, $m_p$ is the mass of the proton,
$\mu=0.6$ is the mean molecular weight of a plasma with primordial
abundances, and $G$ is the gravitational constant. We can define the
core radius of a cluster as $R_c=R_{vir}/p$, and arbitrarily set $p=15$ for 
all the clusters.  Using these
quantities and assuming the $\beta$-profile, we compute the profile of
the $y$ parameter (Eq. \ref{eq:yparam}) associated with each cluster.  
 The cluster atmosphere could depart from the spherical assumption made in
our study. The most extreme geometrical variation would occur if the cluster
is oblate or prolate with its unique axis oriented along the line of sight.
Assuming typical ellipticities (see discussion in \cite{birkinshaw91} and 
\cite{hughes98} and references therein), the Compton parameter could be, at 
the very most, multiplied by a factor 0.5 to 2 due to the cluster
asphericity. The second physical assumption we make concerns the intra-cluster
gas distribution. There is still no strong evidence of a temperature decrease 
with radius (at large radii) from recent X-ray observations (apart from 
cooling flow and merger clusters). Chandra results, with large error bars,
are consistent with a slightly decreasing temperature profile 
\cite[]{markevitch2000} whereas XMM-Newton higher precision observations show 
a ``flat'' profile 
consistent with the isothermal assumption \cite[]{arnaud2001} at least up to
$0.7\,R_{vir}$. Therefore in our study, we make the conservative assumption
of isothermality.
\par\medskip
The kinetic SZ anisotropy map is obtained from the thermal SZ 
map by introducing the radial component of the cluster peculiar velocity.
In the assumption of an isotropic Gaussian distribution of the 
initial density perturbations, the initial power spectrum $P(k)$ gives a 
complete description of the velocity field through the three--dimensional 
{\it rms} velocity $\sigma_v$ predicted by the 
linear gravitational instability at a scale $R$. This velocity is given by:
\begin{equation}
\sigma_v=a(t)\,H(t)\,f(\Omega_m,\Lambda)\left[\frac{1}{2\pi^2}\int^{\infty}_0 P(k)
W^2(kR)\,dk\right]^{1/2}
\label{vrms:eq}
\end{equation}
where $a(t)$ and $H(t)$ are respectively the expansion parameter and the 
Hubble constant. $W$ is the Fourier transform of the window function over 
which the variance is smoothed. The function 
$f(\Omega_m,\Lambda)$ can always be approximated by 
$f(\Omega_m,\Lambda)=\Omega_m^{0.6}$ \cite[]{peebles80,lahav91}.
Furthermore in the assumption of linear regime and a Gaussian distribution 
of the density fluctuations, the structures move with respect to the 
Hubble flow with peculiar velocities, $v$, following a Gaussian distribution
$f(v)=\frac{1}{\sigma_v\sqrt{2\pi}}\exp(\frac{-v^2}{2\sigma_v^2})$ which is 
fully described by $\sigma_v$. This prediction is in agreement with numerical 
simulations \cite[]{bahcall94,moscardini96}.
We therefore compute the three--dimensional 
{\it rms} peculiar velocity using Eq. (\ref{vrms:eq}) for our three
cosmological models. At $z\simeq 0$, we find it ranges between 400 and 500 
km/s (Fig. \ref{fig:vsim}), in good agreement with the observed velocity 
interval $300<v_{rms}<700$ 
km/s \cite[]{hudson94,giovanelli96,moscardini96}. Our values, especially
at low redshifts, could be underestimated by a factor up to 40\% due to the
cluster non-linear evolution at late-time growth \cite[]{colberg2000}.
Nevertheless, we do not correct for it. We will see in the next section 
(Fig. \ref{fig:vszk}) that the expected effect will remain negligible compared
to the other contributions to the error in the peculiar velocity.  
Assuming a random distribution of angles, the amplitude of the
peculiar velocity for individual clusters is drawn at random from specified
the Gaussian distribution, and it is then assigned to the clusters one by one.
\section{The error in individual cluster peculiar velocities} \label{sec:meth}
\subsection{Method}
The radial component of the galaxy cluster peculiar velocity can be
determined using the ratio of thermal to kinetic SZ effect combined
with the intra-cluster temperature. This requires a good separation of the
thermal and kinetic components of the SZ effect which can be achieved, in
principle, through accurate measurements at two or more 
millimetre/submillimetre wavelengths \cite[]{hobson98,bouchet99}
as expected for the Planck mission.\par
Astrophysical or instrumental contributions will introduce systematic 
errors into the determination of the peculiar velocity $v_r$. In this context,
we examine the contributions of the background population of galaxy 
clusters, and then that of both the SZ clusters, and the CMB primary 
anisotropies. The results are given for three
cosmological models using simulated maps convolved with a Gaussian beam 
of 5 arcmin (the Planck effective resolution for
SZ measurements).
\par
The CMB primary temperature fluctuations dominate down to angular scales of 
about 5 arcmin with amplitudes varying with scale and cosmological model.
More generally, each contributing source to the signal (astrophysical or 
instrumental) has its own particular distribution in the amplitude-scale 
space. Therefore, the measurement of the SZ effect at a cluster scale (a 
few arcminutes to a few degrees) picks up a spurious signal from all the
contributions at scales up to the cluster scale (typically the 
angular scale corresponding to its virial radius). \\
Following \cite{aghanim97}, our method consists of detecting the cluster 
through its thermal SZ effect
and measuring the kinetic SZ effect ($\delta T/T$) at this position. The
accuracy of the peculiar velocity determination depends on the accuracy
of both of these measurements. We compute the error to the 
peculiar velocity due to the $\delta T/T$ term using a spatial filter.
The filter is characterised by two windows: a central disk 
(centred on the cluster position) and an external ring, and thus by three 
parameters (the radius of the central 
disc, and the inner and outer radii of the ring). 
The optimum filter is chosen, through the thermal SZ effect measurement,
using only the data i.e. for the Planck resolution, 
the observed cluster profile (convolved with the Planck beam). In practice, 
due to beam dilution effects, we have
obtained an optimum spatial filter that can be applied to a wide range
of clusters. In this case, the disc corresponds to the region where the
Compton parameter is greater than 70\% of the observed central value, and 
the ring is defined by its inner radius $0.5\times\mbox{FWHM}$ and its width 
$\Delta R=2$ pixels.\par
\begin{figure}
\epsfxsize=\columnwidth
\hbox{\epsffile{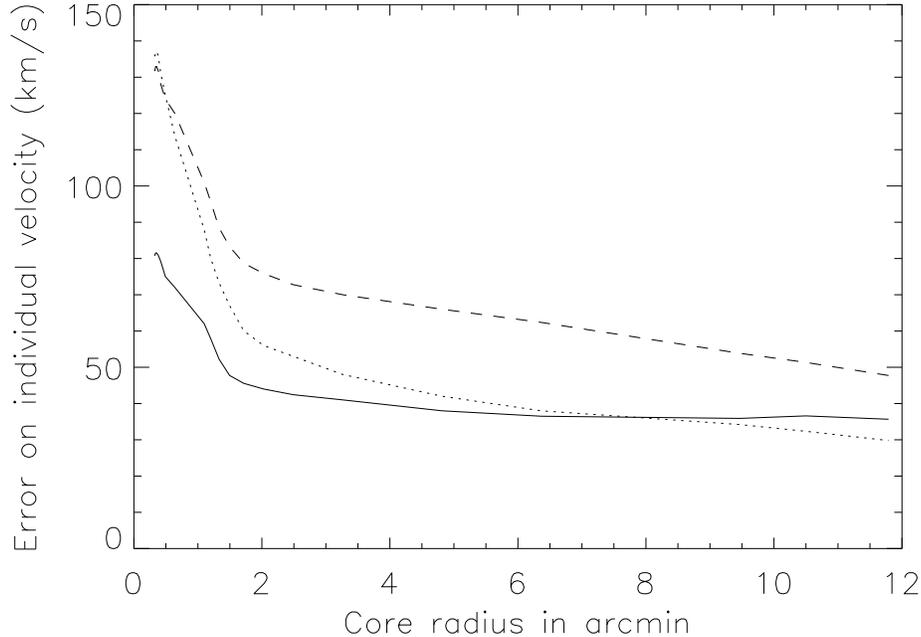}}
\caption{{\small\it The {\it rms} error in the peculiar velocity due
to the kinetic SZ secondary anisotropies of galaxy
cluster background population. The error is given for an individual 
galaxy cluster with $y_0\simeq 10^{-4}$ as a function of its core radius in 
arcmin and for three cosmological
models. The solid, dotted and dashed lines represent respectively the
standard model ($\Omega_m=1$, $\Omega_{\Lambda}=0$), the open model 
($\Omega_m=0.3$) and the flat model with a cosmological
constant ($\Omega_m=0.3$, $\Omega_{\Lambda}=0.7$).}}
\label{fig:vszk}
\end{figure}
\begin{figure}
\epsfxsize=\columnwidth
\hbox{\epsffile{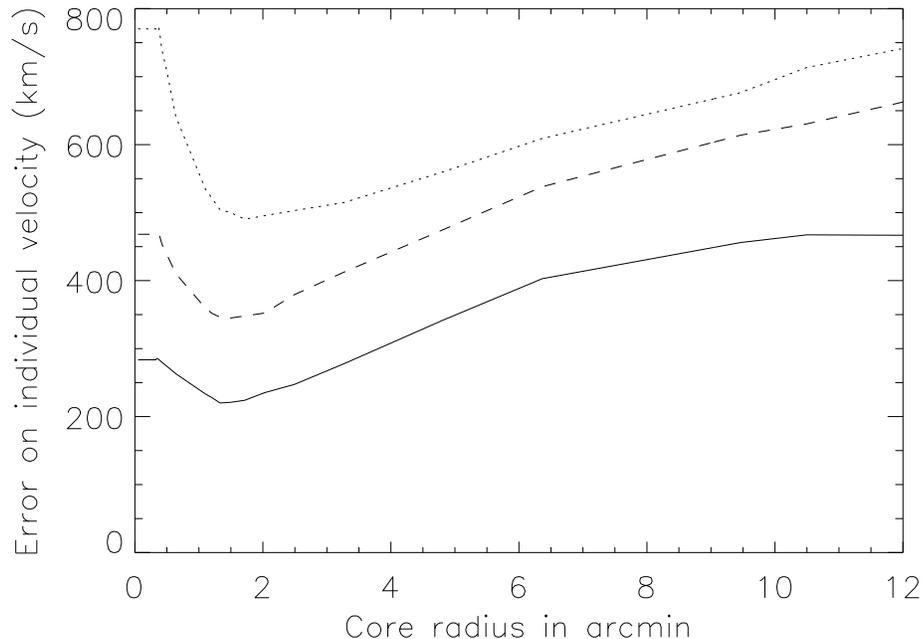}}
\caption{{\small\it For a $y_0\simeq 10^{-4}$ individual cluster with a 
core radius varying
between 0 and 12 arcmin, we plot the {\it rms} error in the radial component
of the peculiar 
velocity due to the kinetic SZ secondary anisotropies of the galaxy
cluster background population plus the CMB primary anisotropies. 
The line-styles represent the same cosmological models as 
in Fig. \ref{fig:vszk}. }}
\label{fig:vtot}
\end{figure}

The peculiar velocity of a targeted cluster is then obtained by computing the
difference of the thermal and kinetic components of the SZ
effect, between the two windows of the filter. Consequently, on a 
``blank field'' containing foreground or
background contributions (CMB, or cluster population, or galactic residuals),
the filter is used to measure the expected contamination due to some, 
or all of these contributions, to the $\delta T/T$ term which is expressed 
in terms of an error $\delta v_r$ on the peculiar velocity. 
In our case and due to the beam dilution, the contribution to the error in 
the peculiar velocity associated with the
amplitude of the thermal SZ effect just acts as a normalisation factor
$\delta v_{rms}\propto 1/y$. In all the following results shown in Figs.
\ref{fig:vszk}, \ref{fig:vtot}, \ref{fig:vfor} and \ref{fig:vft}, the errors 
are computed for a cluster whose central Compton 
parameter is $y_0\simeq 10^{-4}$.\par
To evaluate the accuracy of the peculiar velocity determination, we use
the optimum spatial filter. We compute the differential signal in the 
filter windows and infer the induced error for individual clusters 
$\delta v_r$ using many random 
positions on the simulated maps, convolved by a 5 arcmin beam. 
We obtain errors 300 
$\delta v_r$ in different positions from which we derive the {\it rms}
error in the velocity, $\delta v_{rms}$. This operation is repeated for
different cluster sizes (expressed in terms of the core radius) as we 
expect that the spurious contributions vary with the angular scale. 
\subsection{Results}
We first analyse simulated maps of the background fluctuations induced by the
kinetic SZ effect of a synthetic population of galaxy clusters. 
In order to separately estimate the contribution to the error from the
background SZ kinetic fluctuations no other contributing source
is taken into account at this stage of the analysis. We plot, in 
Fig. \ref{fig:vszk}, the obtained 
$\delta v_{rms}$ as a function of the cluster core radius for three 
cosmological models.  
As a general trend for all the cosmological models, the {\it rms}
error in the individual peculiar velocity decreases rather abruptly with 
increasing core radius for
clusters with core radii up to 2 arcmin; whereas for larger radii
the uncertainty decreases very slowly. This variation of the {\it rms} 
error with cluster size can be understood by the fact that, for extended
clusters (i.e., large core radii), the temperature fluctuations (around zero) 
induced by the kinetic SZ effect of the background clusters are averaged
out. Whereas for small core radii, the effect of beam dilution dominates.
When we compare the error as a function of the cosmological 
model, we note that $\delta v_{rms}$ is the lowest for the standard
model with no cosmological constant ($\Omega_m=1$, solid line). It ranges 
between 
less than 90 km/s (at maximum) and about 40 km/s for the largest core radii.
In the open model ($\Omega_m=0.3$, dotted line), the error reaches about 150 
km/s for very small core radii, and can be as low as about 30 km/s for
extended clusters. This
is due to the combined effect of the cluster higher number counts 
in an open model, especially at high redshifts, and the fact that 
the velocity dispersions are higher in an open model. In the flat model with a
non-zero cosmological constant ($\Omega_m=0.3$, $\Omega_\Lambda=0.7$, 
dashed line),
the cluster peculiar velocities and the number of sources are larger than 
in the other models.  As a consequence of the late-time non-linear growth 
of the galaxy clusters, the predicted error due to the kinetic SZ background
population could be, roughly, up to 40\% larger than the quoted numbers.
\par\bigskip
In the second step, we analyse a simulated $\delta T/T$ map containing
both the primary CMB anisotropies and the kinetic SZ fluctuations due to the 
population of galaxy clusters studied above. The first thing we note in
comparing Figs. \ref{fig:vszk} and \ref{fig:vtot}, is that the
primary CMB dominates the uncertainties, it induces larger errors than
the SZ kinetic population and it has a different behaviour.
In our three cosmological models, unlike the SZ
background contribution, the CMB contribution increases in amplitude with 
increasing cluster size. In fact, $\delta v_{rms}$ decreases (beam dilution
effect) and then increases with increasing core radius with a minimum at about 
1.5 arcmin. This arises from the fact that the amplitude of the primary 
temperature fluctuations increase with the angular scale in 
the range sampled by our spatial filter (acoustic peak). Consequently, this 
larger 
contribution to $\delta T/T$ induces larger errors in the velocity at large
core radii. The open model (dotted
line) exhibits the largest errors which reach about 800 km/s, with a 
minimum of about 500 km/s near 1.5 arcmin. In the
flat model with a non-zero cosmological constant (dashed line), the 
errors culminate at about 650 km/s for extended clusters
their minimum is about 350 km/s near 2 arcmin and they reach about
450 km/s for small core radii. The standard model ($\Omega_m=1$, solid line),
predicts the smallest error with
a maximum, for extended clusters, of about 450 km/s a minimum at 200 km/s and
an error of about 300 km/s for the smallest core radii.  In this context,
the contribution from the non-linear evolution of clusters is negligible with
respect to the CMB primary contribution. The linear approximation used will 
thus almost not affect our results and conclusions.
\par
\begin{figure}
\epsfxsize=\columnwidth
\hbox{\epsffile{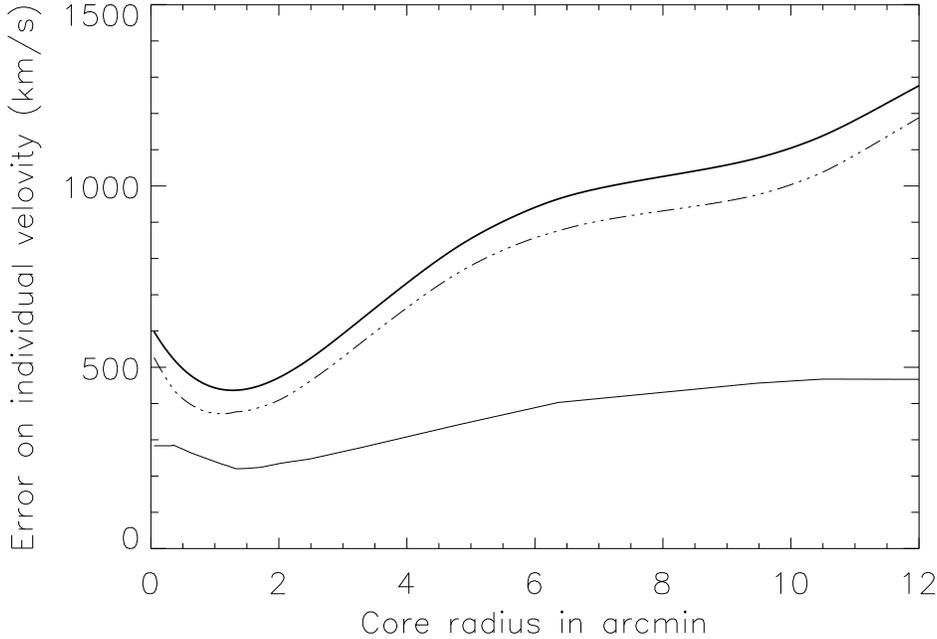}}
\caption{{\small\it The dotted long-dashed line 
represents the contribution to the rms error in the peculiar velocity due 
to the astrophysical residuals from the component separation and to the 
Planck-like instrumental noise (other than CMB and SZ). The thin solid line 
stands for the contribution of CMB and SZ kinetic background population in 
an $\Omega_m=1$ model. The thick solid line represents the total rms errors.}}
\label{fig:vfor}
\end{figure}

\begin{figure}
\epsfxsize=\columnwidth
\hbox{\epsffile{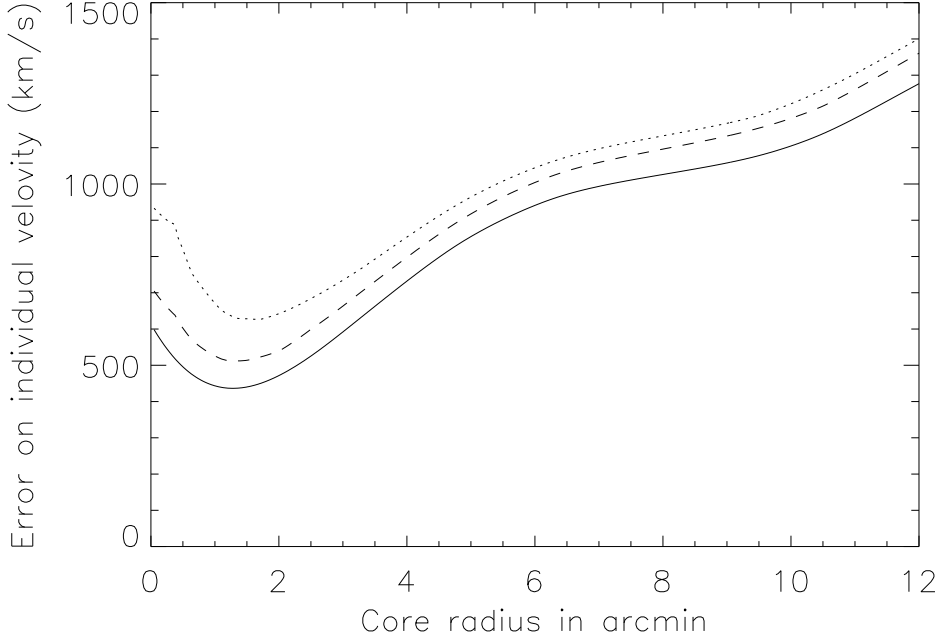}}
\caption{{\small\it The total rms error in the individual velocity due 
to all sources of confusion (CMB, background kinetic SZ, galactic 
residuals, residuals of component separation, Planck-like instrumental 
noise). The line-styles stand for the same cosmological results as in
Fig. \ref{fig:vszk}.}}
\label{fig:vft}
\end{figure}

\bigskip
So far, we have studied two cosmological signals (CMB and kinetic SZ)
that contribute as sources of error in the velocity determination, and 
that are not separated (they have the same spectral signatures). In 
practice, the thermal SZ maps contain information on the position and 
shape of galaxy clusters that helps us in measuring the peculiar velocities.
We have used this information in very simple way by associating the position
of the maximum detected thermal SZ signal with the central position of the
cluster, and also by optimising the spatial filter using the observed thermal
map. More sofisticated methods have been proposed to use the additional phase
information of the different signal in order to reduce the error due to the 
CMB fluctuations \cite[]{scherrer91,naselsky2000}. This kind of method, 
applied for the best case of 
point-like sources \cite[]{naselsky2000}, has shown its usefulness in 
detecting and extracting the point sources buried in a CMB signal. 
The efficiency of such a method, in reducing the CMB contribution to the 
peculiar
velocity, for resolved sources convolved with the instrumental beam (as in
our case) remains to be studied. In addition, our 
analysis has been done in an ideal way because we did not take into account 
the fact that a $\delta T/T$ map will result from a component separation.
In this context, we expect that the residuals from the component separation
will contribute to the error in the peculiar velocity. Among the residuals, an 
important 
contribution comes from the instrumental noise, which in the case of Planck,
is about $2\times 10^{-6}$ {\it rms} $\delta T/T$. For a standard model with 
$\Omega_m=1$, we derive, from a previous study \cite[]{aghanim97}, the 
{\it rms} error due to the instrumental noise and the residual signals from 
the component separation (Fig. \ref{fig:vfor}, dotted long-dashed line).
The component separation was performed using Wiener filtering 
\cite[]{bouchet99}, and the residuals account for the galactic contributions
and the instrumental noise. The total $\delta v_{rms}$ (including the 
contribution from the CMB and kinetic SZ effect (Fig. \ref{fig:vtot}, 
solid line)) is represented by the thick solid line in Fig. \ref{fig:vfor}. 
This error in the peculiar velocity due to the residuals remaining after the 
component separation and to instrumental noise, is thus mostly independent 
of the cosmological 
model, and it should be present in the Planck measurements of
the peculiar velocities using the SZ effect at this level, independent of 
the cosmological model. Furthermore, the induced error is dominant for the 
standard model, and dominant or comparable to the CMB+SZ error in the other
cosmological models. In this context, the phase information will not
help and we do not expect a reduction of the total error in the velocity.
The resulting total {\it rms} errors on the
velocity of individual clusters are displayed in Fig. 4 where the solid 
line represents the standard ($\Omega_m=1$ model), and the dotted and 
dashed lines stand respectively for the open and flat models. 
\section{Bulk velocities}
The error in the individual velocities of galaxy clusters are rather 
large due to several contributions. However, we can obtain
meaningful and valuable information on the velocity fields through
statistical analyses.  
Over large scales, one accessible piece of statistical information derived
from the cluster peculiar velocity is the bulk velocity. It is defined
as the centre-of-mass velocity of a specified region, and it is given by the
integral of the peculiar velocities over a selected volume specified by a
selection function. 
In a given volume containing $N$ clusters with individual peculiar 
velocities $v_i$ each measured with an accuracy $\sigma_i$, the best estimate 
of the bulk velocity $V_{bulk}$ is the mean weighted velocity: 
\begin{equation}
V_{bulk}=\frac{\displaystyle\sum_{i=0}^N\frac{1}{\sigma_i^2}v_i}{\displaystyle 
\sum_{i=0}^N\frac{1}{\sigma_i^2}}.
\label{eq:vb}
\end{equation}
We compute the error due to the astrophysical and instrumental contributions,
in the bulk velocity, using the following relation:
\begin{equation}
\sigma_{bulk}^2=\frac{1}{\displaystyle\sum_i \frac{N_i}{\sigma_i^2}},
\label{eq:vb}
\end{equation}
where $N_i$ is the number of clusters of mass $M_i$ and accuracy $\sigma_i$
at a given redshift.
\par
In order to avoid local non-linear or correlation effects, we focus on
scales that are large enough to allow a significant measurement of the
bulk velocity. We choose an illustrative 100$h^{-1}$ Mpc scale and 
investigate, for redshifts between 0 and 2, the accuracy of the bulk 
velocities when we take into account the major sources contributing to
the error (CMB primary fluctuations, kinetic 
SZ fluctuations of a background cluster population, residuals
due to component separation process and instrumental noise).
In each ``box'' of 100$h^{-1}$ Mpc size defined by its $\Delta \Omega$ and
$\Delta z$ at redshift $z$, we use the PS number counts to 
compute the predicted number of clusters.
\begin{figure}
\epsfxsize=\columnwidth
\hbox{\epsffile{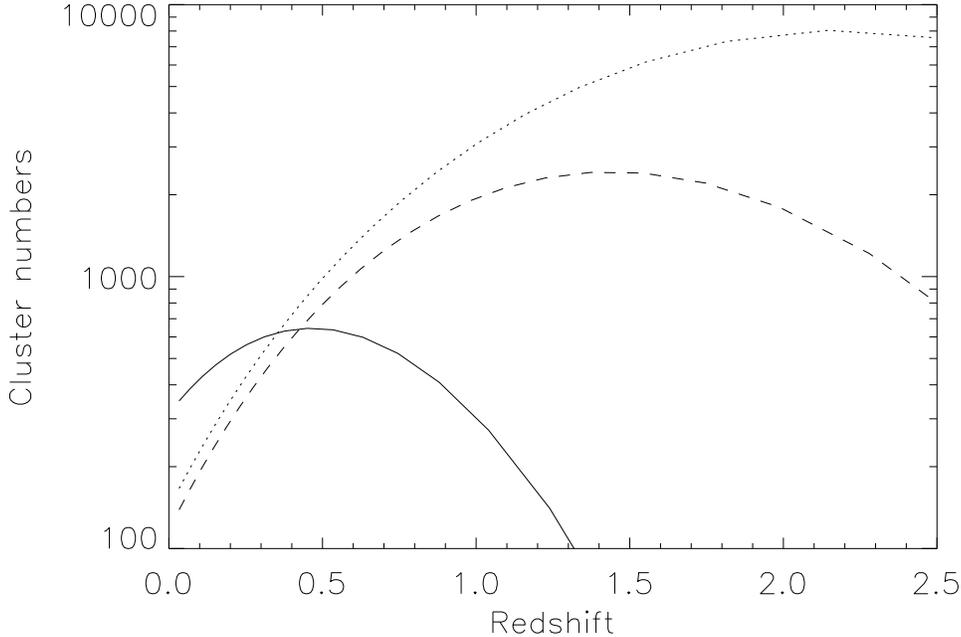}}
\caption{{\small\it The number of clusters as a function of
redshift in the 100$h^{-1}$ Mpc size boxes. The solid line represents the 
standard model ($\Omega_m=1$), the dotted line represents the open (low
$\Omega_m$) model and finally the dashed line shows the results for a 
flat model with a non-zero cosmological constant.}}
\label{fig:numc}
\end{figure}
For each of these clusters, we
compute the core radius $\theta_c$ and the central Compton parameter 
$y_0$. Thus using the results displayed in Fig. \ref{fig:vft}, we can 
associate with each cluster 
in a given volume an error in its individual peculiar velocity. 
Finally using Eq. (\ref{eq:vb}), we compute the overall accuracy 
$\sigma_{bulk}$ in each 100$h^{-1}$ Mpc 
typical size volume and show the results as a function of redshift in 
Fig. \ref{fig:vbulk}. \par
\begin{figure}
\epsfxsize=\columnwidth
\hbox{\epsffile{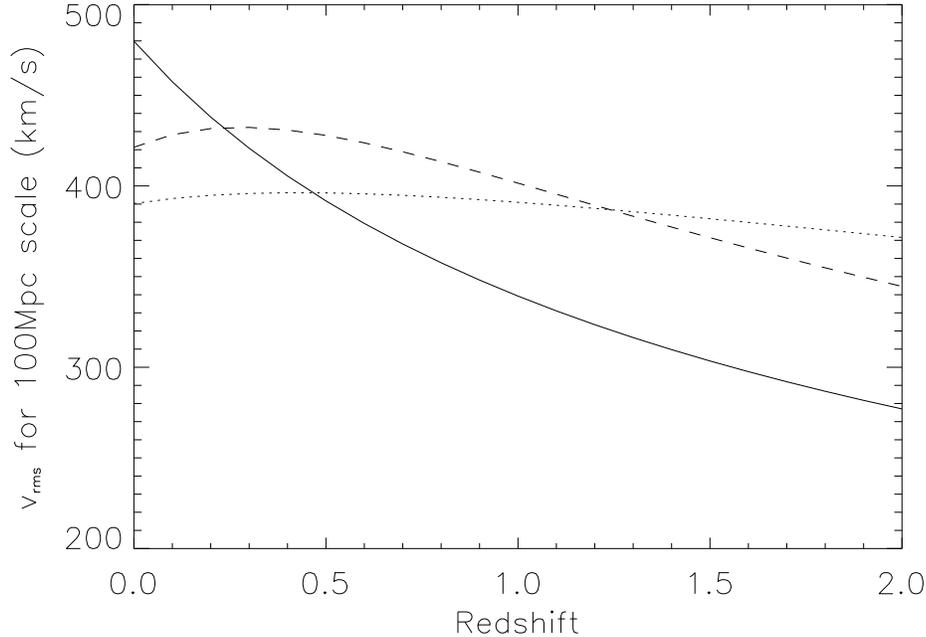}}
\caption{{\small\it The {\it rms} peculiar velocity for a 100 Mpc scale, in 
linear theory, as a function of redshift for three cosmological models. 
The solid, dotted and dashed lines represent respectively the
standard model, the open model and the flat model with a cosmological
constant.}}
\label{fig:vsim}
\end{figure}
\begin{figure}
\epsfxsize=\columnwidth
\hbox{\epsffile{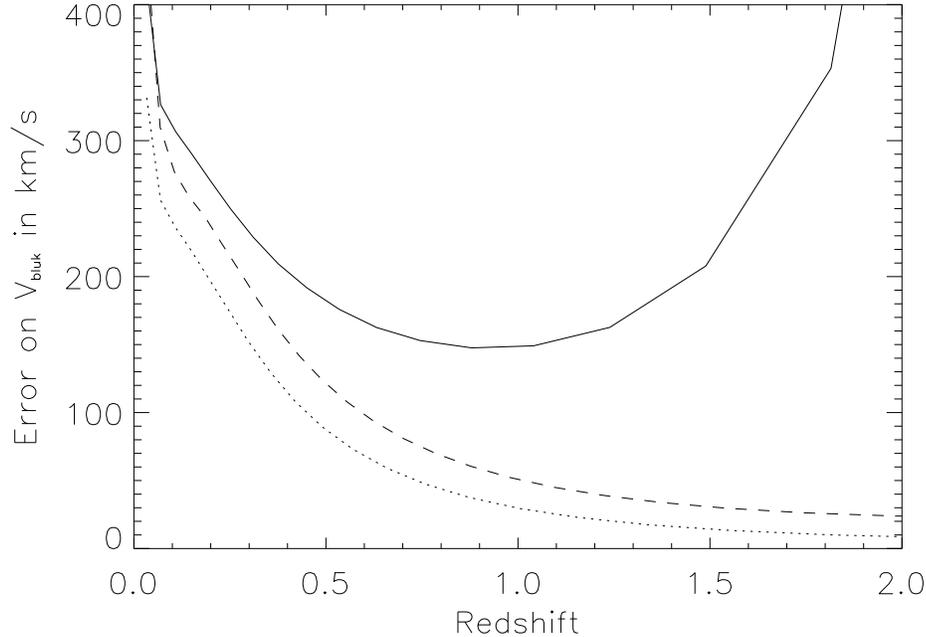}}
\caption{{\small\it The rms error in the bulk velocities measured in 
volumes of 100$h^{-1}$ Mpc typical size. The solid line represents the 
standard model ($\Omega_m=1$), the dotted line represents the open (low
$\Omega_m$) model and finally the dashed line shows the results for a 
flat model with a non-zero cosmological constant.}}
\label{fig:vbulk}
\end{figure}

In the standard model, the bulk velocity at the 100$h^{-1}$ Mpc scale can 
be affected by an error as large as 400 km/s at small redshifts. This
error decreases with redshift down to about 150 km/s by $z=0.8$. It 
remains almost constant until $z\simeq 1.3$ and increases again to very
large values by $z=2$. This increase is associated with the lack of clusters
at high redshifts in high $\Omega_m$ models (see Fig. 
\ref{fig:numc}). In an open model the number of clusters being much higher,
the accuracy in the bulk velocity is better than in the standard case. The
overall error decreases very rapidly. It is of the order of 80 km/s at $z=0.5$ 
and reaches 30 km/s at $z=1$. At $z=2$, the accuracy is as small as 10
km/s. It increases at high redshift, again when the number of clusters
decreases. Similarly in the flat model with a non-zero 
cosmological constant, the error in the bulk velocity decreases rapidly with 
redshift; but it is slightly larger than in the open case at all redshifts. 
The error ranges between 400 
km/s (at low redshifts) and 25 km/s at $z=2$. At $z=0.5$ and $z=1$, it reaches 
respectively 100 and 50 km/s. In the low density models, we thus expect 
to achieve very accurate 
measurements of the bulk motion using the SZ effect. This can be illustrated
by comparing a rough estimate of $V_{bulk}$ for a 100 Mpc scale (Fig. 
\ref{fig:vsim}), computed in the linear theory (Eq. \ref{vrms:eq}), with the
estimated error (Fig. \ref{fig:vbulk}). On the contrary, the precision for 
the standard model is rather poor ($\simeq 2\sigma$ around $z=1$).
\par
\begin{figure}
\epsfxsize=\columnwidth
\hbox{\epsffile{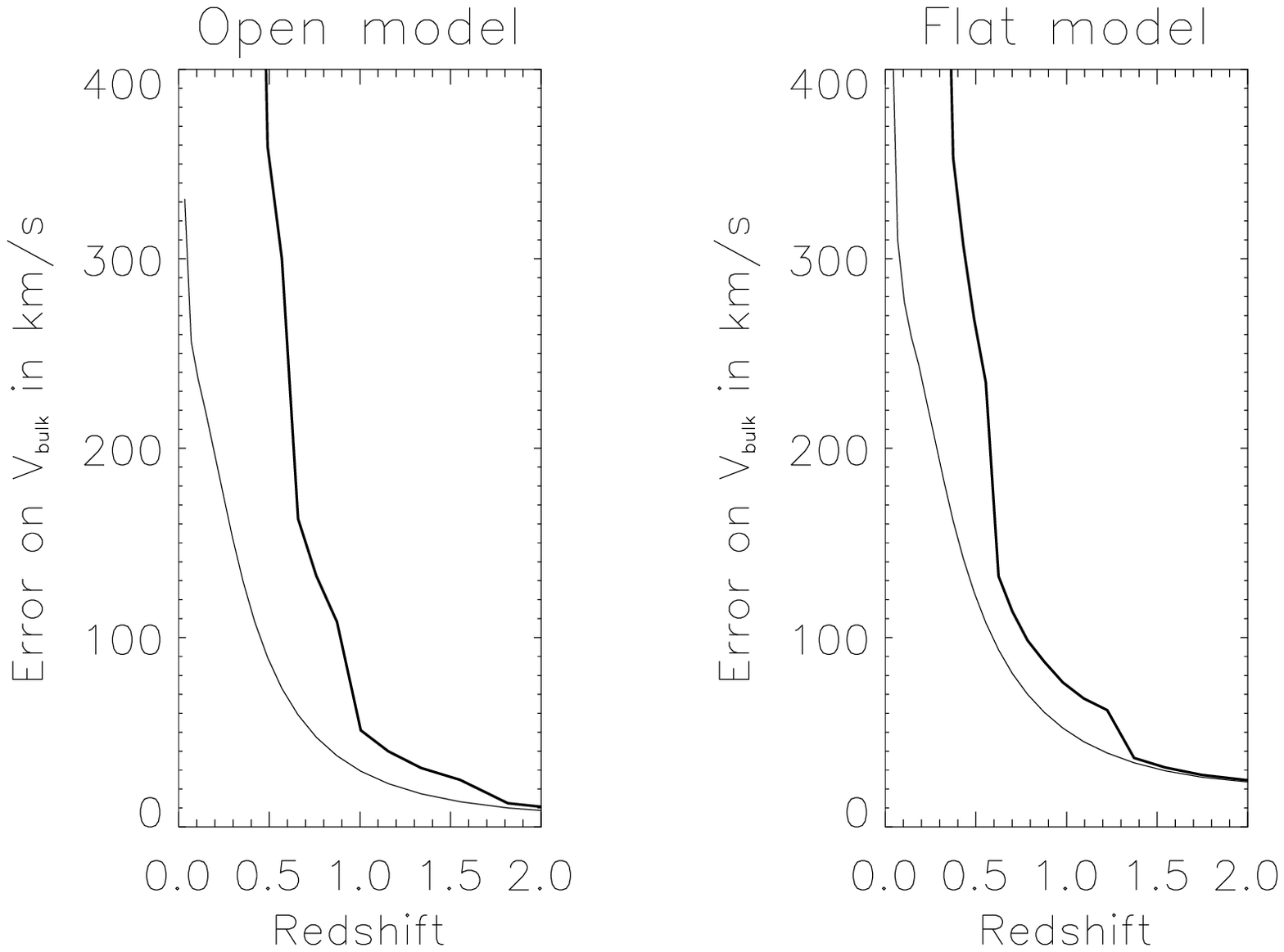}}
\caption{{\small\it The rms error in the bulk velocities 100$h^{-1}$ Mpc 
typical scale. The solid lines represent the errors when the limitation
due to source confusion is taken into account. The thin lines shows the 
results uncorrected form the confusion effects (open model is in
the left panel and flat model, $\Omega_\Lambda=0.7$, is in the right
panel). The confusion limit is set so that there is
one cluster in 30 independent 5 arcmin beams.}}
\label{fig:vbconf}
\end{figure}

These very impressive accuracies in the bulk velocity measurements would be 
those obtained on 100$h^{-1}$ Mpc scale using the SZ effect and assuming that 
all the SZ sources are detected and used for the computation. This would be 
the case if the sky were to be surveyed
with a very high angular resolution (typically better than 1 arcmin) as
will be achieved by FIRST-Herschel. In the Planck configuration,
for a 5 arcmin beam convolution, source confusion will be the limiting 
factor in cluster detection and will lead 
to a degradation of the bulk velocity accuracy. Source confusion 
affects the measurements in all the cosmological models, but the problem is 
especially severe in low density matter models for which the clusters 
are more numerous at $z>0.3$ (e.g. \cite[]{barbosa96}), as illustrated
in Fig. \ref{fig:numc}. We estimate the effect of source confusion 
on the bulk velocity measurement, and illustrate the results for the low 
density matter models. To do so, we use the canonical confusion limit
for Euclidean number counts. This implies the presence of one source in 30 
independent beams (5 arcmin beam in the Planck configuration). Therefore, the 
total number of clusters on the sky must not exceed $1.9\times 10^5$. This 
condition thus defines,
for our three cosmological models, a detection (or confusion) limit 
$Y_{lim}$ in terms of the integrated Compton parameter above which the 
clusters are detected. For the flat model with non-zero cosmological constant, 
we find that $Y_{lim}\sim 14\times 10^{-3}$ arcmin$^2$. As expected, 
the limit is higher for the open model for which it is about 
$23\times 10^{-3}$ arcmin$^2$. This comes from the fact that more clusters 
are predicted, and the source confusion effect is thus more important.
Taking into account this additional condition in order to estimate
the number of detected clusters, we re-evaluate the error in the bulk 
velocity in the 100$h^{-1}$ Mpc typical size boxes and we compare the results
to those obtained without the correction for source confusion. The results
illustrated for the two low density models are displayed in Fig. 
\ref{fig:vbconf}. The thick lines represent the errors when the limitation
due to source confusion is taken into account as compared with the previous
results (thin lines). In the open model (left panel), the limitation due to 
confusion results in a larger error, of the order of  300 km/s at 
$z=0.5$ and 50 km/s at $z=1$. This is, respectively, almost four
times and twice as large as the previous values.
In the flat model (right panel), the confusion effect increases the error
on the bulk velocity by a factor of about two at $z=0.5$ where the error is
about 260 km/s. At redshift $z=1$, the error is of the order of 70 km/s 
(a factor 1.4 larger than the previous accuracy Fig. \ref{fig:vbulk}).
\section{Discussion and conclusion}
The kinetic SZ effect can be used as an alternative powerful tool to
measure the radial component of the peculiar cluster velocities, and
consequently to trace the matter distribution. Nevertheless, the 
accuracy for individual clusters is limited by the contributions from other 
astrophysical components and instrumental noise, and by the source (or spatial)
confusion with other clusters. In the present study, we 
investigate the limitation of this method. More precisely, we estimate
the {\it rms} error by which the source confusion and the major contributing
sources (CMB primary anisotropies, kinetic SZ due to the
background cluster population, and  noise + residuals
from component separation) can affect the velocity measurement.
\par
The amplitude of the spurious cosmological contributions (CMB and SZ kinetic)
and the source confusion varies with the cosmological model. Therefore
keeping the baryon density constant, we have investigated
three cosmological models, two flat models with and without a cosmological 
constant ($\Omega_\Lambda=0$ and 0.7) and one open model $\Omega_m=0.3$. A 
previous study by \cite{haehnelt96} estimated
the effect of changing the baryon fraction of the universe on the error 
in the individual peculiar velocity.
\par
The contribution of the kinetic SZ background clusters is not a major source
of error (relative to the other contributions). It
represents, depending on the cosmological model, an {\it rms} error
between about 140 and 80 km/s for small cluster sizes, and between 
50 and 30 km/s for more extended clusters. These errors could be up to
40\% larger due to the non-linear growth of the structures, but this still 
does not significantly affect our final results because the total error is 
clearly dominated by the other contributions. Departures from our physical
assumptions, such as possible asphericity or non-isothermality, can affect
the previous values. For an individual cluster the asphericity can modify
its SZ kinetic contribution by a factor 0.5 to 2. However, this is the extreme
variation for the case where the principal axis is aligned along the line of 
sight. In addition for a cluster population with random orientations of the 
ellipticities, we expect that
the net geometrical effect will average out so that the error due
to the background cluster population will not be significantly affected. 
Recent X-ray observations of galaxy clusters show isothermal profiles but 
still leave room for decreasing temperature profiles above $0.7\,R_{vir}$. 
In this case, the contribution of an individual cluster is smaller especially
in the cluster outskirts. As a result, the contribution to the temperature 
fluctuations due to the kinetic SZ effect of the whole cluster population 
should be smaller than in the isothermal assumption. Consequently, the 
departures from the simplest approximations used in our study are likely to
average out, and therefore leave the results almost unchanged.
The {\it rms} error due to 
the CMB primary anisotropies is dominant, compared to SZ background 
clusters, and varies with the cosmological model and the
cluster size. However, it shows the same shape in all models, i.e., large 
velocities 
at small cluster sizes (due to beam dilution effects) which sharply decrease 
and reach a minimum at about 2 
arcmin core radius, followed by an increase at larger scales associated with
the contribution from the acoustic peak. The 
{\it rms} error reaches its maximum value (almost 800 km/s) at small 
cluster sizes in the 
open model, and its minimum value (almost 200 km/s) at about 2
arcmin core radius in the standard model. The last spurious component we 
have taken into 
account is that due to the residual signals from a component separation
including Planck like instrumental noise. This contribution is of the order
of, or dominates, the CMB depending on the amplitude of the cosmological 
``errors'' (that is depending on the cosmological model). The 
induced {\it rms} error shows a maximum value for large cluster sizes, where 
it reaches about 1000 km/s, it decreases to 400 km/s at 2 arcmin core radius 
then increases again up to 500 km/s. The contribution from residuals in the
component separation could well be smaller if we use better-adapted component 
separation techniques, such as the proposed maximum entropy method 
\cite[]{hobson98}. All the resulting accuracies are obtained for a rather 
important thermal effect ($y_0=10^{-4}$). The velocities vary as a function 
of $1/y_0$, and ``weaker'' clusters are affected by larger errors. Another 
source of error in the peculiar 
velocity is the relativistic correction to the SZ effect, which is not taken 
into account in our study. It has been 
evaluated by \cite{holzapfel97} who found it to be of the order of 
360$(kT_e/10 \mbox{keV})^2$ km/s. Furthermore, our results do not take 
into account the 
error due to the uncertainty in the measured intra-cluster temperature. 
As previously mentioned, the recent X-ray
satellite observations will allow very precise temperature measurements with 
an accuracy of 5 to 10\%. \par\medskip
The measurement of the radial 
component of the peculiar cluster velocity using the SZ effect 
is only marginally possible for the ``strongest'' clusters, and then only in a 
rather narrow spatial window corresponding to clusters with core radii 
around 2 arcmin. The number of clusters matching the two criteria (size 
and amplitude) over the sky is small. Therefore, we rather turn to a 
statistical approach. The velocity dispersion is not the appropriate
quantity to study as it is very sensitive to systematic 
effect contributions. The bulk velocity, in turn, is a better-adapted 
statistical quantity which 
can be derived from the SZ measurement of the individual peculiar velocities.
Based on the all-sky high sensitivity
SZ survey which will be provided by the Planck mission, we propose
a strategy to trace the velocity field by measuring the bulk flows on very 
large scales up to $z=1$, or greater. The method relies on averaging, over 
large volumes, the peculiar velocities of the individual
clusters detected therein. In this context, we have evaluated 
the {\it rms} error associated with the bulk velocity in 100$h^{-1}$ Mpc 
typical size boxes by generalising, to 
a population of clusters predicted by the PS formalism, the 
results obtained for individual clusters. 
The accuracy in the bulk velocity ($\sigma_{bulk}$) depends on two main 
quantities: the accuracy
of the individual radial velocities ($\sigma_i$) and the number of clusters
($N_i$). The smaller $\sigma_i$ then the smaller is $\sigma_{bulk}$. The 
larger $N_i$ then the smaller is $\sigma_{bulk}$. Therefore we find
a rather poor accuracy in the standard flat model for which the predicted 
number of clusters is small. In contrast, in both low matter density 
models ($\Omega_m < 1$), the accuracy of the bulk velocities is higher.
In our study, we find that the accuracy of the bulk flow determination is 
thus dominated by 
the numerous low mass clusters. However, large cluster numbers result 
in source confusion
which in turn limits our ability to detect individual clusters. It thus 
decreases the number of clusters useful for the bulk velocity estimates which 
results in a larger $\sigma_{bulk}$. Taking into account the fact that source 
confusion will prevent us from detecting all the individual SZ clusters in 
the Planck all-sky survey, we have estimated the new errors on the bulk 
velocities (Fig. \ref{fig:vbconf}). Our method cannot be applied for 100
$h^{-1}$ Mpc boxes, due to lack of sources above the detection limit, below 
a redshift of 0.15 and 0.35 
in, respectively, the flat and open models. In the flat model, the errors are
increased with decreasing redshifts by a factor reaching 3 at $z=0.22$. 
In the open model, the degradation due to source confusion is even stronger. 
The degradation factor is 2.75 at $z=0.66$ increasing to 4 at $z=0.5$. There
is almost no loss of accuracy at $z\ge 1.8$.
In the low density models, the bulk velocities measured with the SZ effect 
exhibit large errors at small redshifts and small errors at high redshifts.
These results show that the the bulk velocities on large scales, especially at 
high redshifts, can be accurately measured and mapped through the SZ effect
(see Figs. \ref{fig:vsim} and \ref{fig:vbconf}).  
This is opposite to the case of classical velocity measurements where the
errors are larger at higher redshifts (e.g. \cite{willick99} and 
references therein). The velocity field could thus be mapped using these two
complementary approaches (``classical'' and SZ) as a function of the redshift
range that is probed. 
\par\bigskip
Assuming a low matter density universe ($\Omega_m=0.3$), a proposed strategy 
to measure the bulk velocities on large scales using the SZ effect would be to 
average the individual peculiar velocities in large boxes between redshifts 
of 0 and 1, or more. 
The size of the boxes would be chosen so that the obtained accuracy in the
bulk velocity is of the order of a few tens km/s. This allows an inhomogeneous 
sampling of the universe. A sampling with 100$h^{-1}$
Mpc typical size boxes gives satisfying accuracy around $z=1$ ($\sigma_{bulk}
\sim 80-50$ km/s), but gives larger errors around $z=0.5$ especially
in an open model. One way of overcoming this problem, would be to 
increase the statistical cluster sample by estimating the bulk velocity on 
larger scales ($>100h^{-1}$ Mpc) at intermediate redshifts. Another
possibility would be to measure the bulk velocity of the local universe up 
to a certain
redshift, typically $z=0.2$ for flat universe and $z=0.4$ for open universe.
In order to measure the velocities using the SZ effect, 
we need, in addition to a measurement of the Compton parameter $y$ and the 
temperature fluctuation of each cluster $\delta T/T$, its
electron intra-cluster temperature $T_e$ and its redshift $z$. At intermediate
redshift ($0<z<0.5$), we could use for a very large fraction of the sky (about
$\pi$ steradian), the SDSS redshift determinations and the available X-ray 
observations (ROSAT, ASCA, XMM-Newton, Chandra) to derive the temperature. 
The temperature could be obtained either through spectroscopic measurements,
or through empirical relations with general properties established on smaller
samples. At higher redshifts ($0.5<z<1$), we could focus on a 
few selected regions of the sky. The regions in this case, should correspond 
to sky selected areas on which multi-wavelength observations will provide 
cluster surveys with their redshifts and temperatures or masses (VIRMOS, XMM,
Chandra, MEGACAM). For a better sky coverage, specific complementary 
observations could be programmed before the Planck mission.
\par
The study of galaxy cluster peculiar velocities is advantageous since,
on scales probed by clusters, the underlying density
fluctuations are largely in the linear regime and therefore very close
to the initial conditions from which large-scale structure developed. The 
large-scale velocity field represents a direct 
prediction of the cosmological model and depends on the power spectrum
and matter density. Following our proposed strategy, based on the all-sky SZ
survey provided by Planck, we will be able to study the velocity field over 
cosmologically important scales and examine its evolution. We could use, as 
proposed by \cite{ferreira99} and \cite{juszkiewicz99}, the evolution of 
the mean relative velocity of pairs of boxes as a function of their separation 
to directly constrain the cosmological parameters. We could also apply
methods based on the
comparison between the reconstructed total matter distribution from velocity 
fields \cite[]{bertschinger89,zaroubi99} and that traced by 
galaxies to constrain the matter density; the constraints on the power 
spectrum over these large scales can be directly compared to CMB constraints.  
\begin{acknowledgements}
The authors would like to thank M. Arnaud, C. Balland and J. Bartlett 
for interesting discussions, and A. Jones for his very careful reading. 
They also thank an anonymous referee for helpful comments.
\end{acknowledgements}


\begin{thebibliography}{{Kashlinsky} \& {Atrio-Barandela}(2000)}

\bibitem[{Aghanim} et~al.(1997)]{aghanim97}
{Aghanim}, N., {De Luca}, A., {Bouchet}, F.~R., {Gispert}, R. \& {Puget}, J.~L.
\newblock 1997, \AaA, 325, 9.

\bibitem[{Arnaud et al.}(2001)]{arnaud2001}
{Arnaud et al.}, M.
\newblock In {\em Clusters of galaxies and the
  high redshift universe observed in X-rays}, Rencontres de Moriond, 2001, Ed.
D.~Neumann, J. T. T.~Van.

\bibitem[{Bahcall} et~al.(1994)]{bahcall94}
{Bahcall}, N.~A., {Cen}, R. \& {Gramann}, M.
\newblock 1994, \ApJ, 430, L13.

\bibitem[{Barbosa} et~al.(1996)]{barbosa96}
{Barbosa}, D., {Bartlett}, J.~G., {Blanchard}, A. \& {Oukbir}, J.
\newblock 1996, \AaA, 314, 13.

\bibitem[{Bertschinger} \& {Dekel}(1989)]{bertschinger89}
{Bertschinger}, E. \& {Dekel}, A.
\newblock 1989, \ApJ, 336, L5.

\bibitem[{Birkinshaw} \& {Gull}(1983)]{birkinshaw83}
{Birkinshaw}, M. \& {Gull}, S.~F.
\newblock 1983, \Natur, 302, 315.

\bibitem[{Birkinshaw} et~al.(1991)]{birkinshaw91}
{Birkinshaw}, M., {Hughes}, J.~P. \& {Arnaud}, K.~A.
\newblock 1991, \ApJ, 379, 466.

\bibitem[{Birkinshaw}(1983)]{birkinshaw83a}
{Birkinshaw}, M.
\newblock In {\em Quasars \& Gravitational Lenses: 24th Liege}, page 134, 1983.

\bibitem[{Bouchet} \& {Gispert}(1999)]{bouchet99}
{Bouchet}, F.~R. \& {Gispert}, R.
\newblock 1999, New Astronomy, 4, 443.

\bibitem[{Bryan} \& {Norman}(1998)]{bryan98}
{Bryan}, G.~L. \& {Norman}, M.~L.
\newblock 1998, \ApJ, 495, 80.

\bibitem[{Carroll} et~al.(1992)]{carroll92}
{Carroll}, S.~M., {Press}, W.~H. \& {Turner}, E.~L.
\newblock 1992, \ARAA, 30, 499.

\bibitem[{Cavaliere} \& {Fusco-Femiano}(1978)]{cavaliere78}
{Cavaliere}, A. \& {Fusco-Femiano}, R.
\newblock 1978, \AaA, 70, 677.

\bibitem[{Challinor} \& {Lasenby}(1998)]{challinor98}
{Challinor}, A. \& {Lasenby}, A.
\newblock 1998, \ApJ, 499, 1.

\bibitem[{Colberg} et~al.(2000)]{colberg2000}
{Colberg}, J.~M., {White}, S. D.~M., {Macfarland}, T.~J., {Jenkins}, A.,
  {Pearce}, F.~R., {Frenk}, C.~S., {Thomas}, P.~A. \& {Couchman}, H. M.~P.
\newblock 2000, \MNRAS, 313, 229.

\bibitem[{Courteau} et~al.(1993)]{courteau93}
{Courteau}, S., {Faber}, S.~M., {Dressler}, A. \& {Willick}, J.~A.
\newblock 1993, \ApJ, 412, L51.

\bibitem[{Dressler} et~al.(1987)]{dressler87}
{Dressler}, A., {Faber}, S.~M., {Burstein}, D., {Davies}, R.~L., {Lynden-Bell},
  D., {Terlevich}, R.~J. \& {Wegner}, G.
\newblock 1987, \ApJ, 313, L37.

\bibitem[{Edge} \& {Stewart}(1991)]{edge91}
{Edge}, A.~C. \& {Stewart}, G.~C.
\newblock 1991, \MNRAS, 252, 414.

\bibitem[{Evrard}(1990)]{evrard90}
{Evrard}, A.~E.
\newblock 1990, \ApJ, 363, 349.

\bibitem[{Faber} \& {Jackson}(1976)]{faber76}
{Faber}, S.~M. \& {Jackson}, R.~E.
\newblock 1976, \ApJ, 204, 668.

\bibitem[{Ferreira} et~al.(1999)]{ferreira99}
{Ferreira}, P.~G., {Juszkiewicz}, R., {Feldman}, H.~A., {Davis}, M. \& {Jaffe},
  A.~H.
\newblock 1999, \ApJ, 515, L1.

\bibitem[{Giovanelli} et~al.(1996)]{giovanelli96}
{Giovanelli}, R., {Haynes}, M.~P., {Wegner}, G., {Da Costa}, L.~N.,
  {Freudling}, W. \& {Salzer}, J.~J.
\newblock 1996, \ApJ, 464, L99.

\bibitem[{Giovanelli} et~al.(1998)]{giovanelli98}
{Giovanelli}, R., {Haynes}, M.~P., {Salzer}, J.~J., {Wegner}, G., {Da Costa},
  L.~N. \& {Freudling}, W.
\newblock 1998, \AJ, 116, 2632.

\bibitem[{Haehnelt} \& {Tegmark}(1996)]{haehnelt96}
{Haehnelt}, M.~G. \& {Tegmark}, M.
\newblock 1996, \MNRAS, 279, 545.

\bibitem[{Hansen} \& {Lilje}(1999)]{hansen99}
{Hansen}, F.~K. \& {Lilje}, P.~B.
\newblock 1999, \MNRAS, 306, 153.

\bibitem[{Hobson} et~al.(1998)]{hobson98}
{Hobson}, M.~P., {Jones}, A.~W., {Lasenby}, A.~N. \& {Bouchet}, F.~R.
\newblock 1998, \MNRAS, 300, 1.

\bibitem[{Holzapfel} et~al.(1997)]{holzapfel97}
{Holzapfel}, W.~L., {Ade}, P. A.~R., {Church}, S.~E., {Mauskopf}, P.~D.,
  {Rephaeli}, Y., {Wilbanks}, T.~M. \& {Lange}, A.~E.
\newblock 1997, \ApJ, 481, 35.

\bibitem[{Hudson} et~al.(1999)]{hudson99}
{Hudson}, M.~J., {Smith}, R.~J., {Lucey}, J.~R., {Schlegel}, D.~J. \& {Davies},
  R.~L.
\newblock 1999, \ApJ, 512, L79.

\bibitem[{Hudson}(1994)]{hudson94}
{Hudson}, M.~J.
\newblock 1994, \MNRAS, 266, 475.

\bibitem[{Hughes} \& {Birkinshaw}(1998)]{hughes98}
{Hughes}, J.~P. \& {Birkinshaw}, M.
\newblock 1998, \ApJ, 501, 1.

\bibitem[{Itoh} et~al.(1998)]{itoh98}
{Itoh}, N., {Kohyama}, Y. \& {Nozawa}, S.
\newblock 1998, \ApJ, 502, 7.

\bibitem[{Jones} \& {Forman}(1984)]{jones84}
{Jones}, C. \& {Forman}, W.
\newblock 1984, \ApJ, 276, 38.

\bibitem[{Juszkiewicz} et~al.(1999)]{juszkiewicz99}
{Juszkiewicz}, R., {Springel}, V. \& {Durrer}, R.
\newblock 1999, \ApJ, 518, L25.

\bibitem[{Kashlinsky} \& {Atrio-Barandela}(2000)]{kashlinsky2000}
{Kashlinsky}, A. \& {Atrio-Barandela}, F.
\newblock 2000, \ApJ, 536, L67.

\bibitem[{King}(1966)]{king66}
{King}, I.~R.
\newblock 1966, \AJ, 71, 64.

\bibitem[{Lahav} et~al.(1991)]{lahav91}
{Lahav}, O., {Rees}, M.~J., {Lilje}, P.~B. \& {Primack}, J.~R.
\newblock 1991, \MNRAS, 251, L128.

\bibitem[{Lamarre} et~al.(1998)]{lamarre98}
{Lamarre}, J.~M., {Giard}, M., {Pointecouteau}, E., {Bernard}, J.~P., {Serra},
  G., {Pajot}, F., {D\'esert}, F.~X., {Ristorcelli}, I., {Torre}, J.~P.,
  {Church}, S., {Coron}, N., {Puget}, J.~L. \& {Bock}, J.~J.
\newblock 1998, \ApJ, 507, L5.

\bibitem[{Markevitch} et~al.(2000)]{markevitch2000}
{Markevitch}, M., {Vikhlinin}, A., {Mazzotta}, P. \& {Vanspeybroeck}, L.
\newblock In astronomy 2000~(Palermo), X, editor, {\em R. Giacconi, L. Stella,
  S. Serio}. APS Conf. Series, 2000.

\bibitem[{Molnar} \& {Birkinshaw}(1999)]{molnar99}
{Molnar}, S.~M. \& {Birkinshaw}, M.
\newblock 1999, \ApJ, 523, 78.

\bibitem[{Moscardini} et~al.(1996)]{moscardini96}
{Moscardini}, L., {Branchini}, E., {Brunozzi}, P.~T., {Borgani}, S., {Plionis},
  M. \& {Coles}, P.
\newblock 1996, \MNRAS, 282, 384.

\bibitem[{Naselsky} et~al.(2000)]{naselsky2000}
{Naselsky}, P., {Novikov}, D. \& {Silk}, J.
\newblock astro-ph/0007133, 2000.

\bibitem[{Nozawa} et~al.(2000)]{nozawa2000}
{Nozawa}, S., {Itoh}, N., {Kawana}, Y. \& {Kohyama}, Y.
\newblock 2000, \ApJ, 536, 31.

\bibitem[{Peebles}(1980)]{peebles80}
{Peebles}, P. J.~E.
\newblock In {\em The large-scale structure of the universe}, 1980.

\bibitem[{Peebles}(1993)]{peebles93}
{Peebles}, P. J.~E.
\newblock In {\em Principles of physical cosmology}, 1993.

\bibitem[{Pointecouteau} et~al.(1998)]{pointecouteau98}
{Pointecouteau}, E., {Giard}, M. \& {Barret}, D.
\newblock 1998, \AaA, 336, 44.

\bibitem[{Press} \& {Schechter}(1974)]{press74}
{Press}, W.~H. \& {Schechter}, P.
\newblock 1974, \ApJ, 187, 425.

\bibitem[{Rephaeli} \& {Lahav}(1991)]{rephaeli91}
{Rephaeli}, Y. \& {Lahav}, O.
\newblock 1991, \ApJ, 372, 21.

\bibitem[{Rephaeli}(1995)]{rephaeli95}
{Rephaeli}, Y.
\newblock 1995, \ARAA, 33, 541.

\bibitem[{Sazonov} \& {Sunyaev}(1998)]{sazonov98}
{Sazonov}, S.~Y. \& {Sunyaev}, R.~A.
\newblock 1998, \ApJ, 508, 1.

\bibitem[{Scherrer} et~al.(1991)]{scherrer91}
{Scherrer}, R.~J., {Melott}, A.~L. \& {Shandarin}, S.~F.
\newblock 1991, \ApJ, 377, 29.

\bibitem[{Sunyaev} \& {Zel'dovich}(1980)]{suniaev80}
{Sunyaev}, R.~A. \& {Zel'dovich}, I.~B.
\newblock 1980, \ARAA, 18, 537.

\bibitem[{Tully} \& {Fisher}(1977)]{tully77}
{Tully}, R.~B. \& {Fisher}, J.~R.
\newblock 1977, \AaA, 54, 661.

\bibitem[{Viana} \& {Liddle}(1996)]{viana96}
{Viana}, P. T.~P. \& {Liddle}, A.~R.
\newblock 1996, \MNRAS, 281, 323.

\bibitem[{Viana} \& {Liddle}(1999)]{viana99}
{Viana}, P. T.~P. \& {Liddle}, A.~R.
\newblock 1999, \MNRAS, 303, 535.

\bibitem[{Walker} et~al.(1991)]{walker91}
{Walker}, T.~P., {Steigman}, G., {Kang}, H.~S., {Schramm}, D.~M. \& {Olive},
  K.~A.
\newblock 1991, \ApJ, 376, 51.

\bibitem[{Willick} et~al.(1996)]{willick96}
{Willick}, J.~A., {Courteau}, S., {Faber}, S.~M., {Burstein}, D., {Dekel}, A.
  \& {Kolatt}, T.
\newblock 1996, \ApJ, 457, 460.

\bibitem[{Willick}(1999)]{willick99}
{Willick}, J.~A.
\newblock 1999, \ApJ, 522, 647.

\bibitem[{Wright}(1979)]{wright79}
{Wright}, E.~L.
\newblock 1979, \ApJ, 232, 348.

\bibitem[{Zaroubi} et~al.(1999)]{zaroubi99}
{Zaroubi}, S., {Hoffman}, Y. \& {Dekel}, A.
\newblock 1999, \ApJ, 520, 413.

\end{thebibliography}
\end{document}